\documentclass[american,aps,prl,showpacs,preprint]{revtex4}
\usepackage[T1]{fontenc}
\usepackage{graphicx}

\makeatletter

\usepackage{babel}
\makeatother
\begin{document}

\title{Complex patterns and tip effect evolution}

\author{Francisco Vera}

\email{fvera@ucv.cl}

\affiliation{Pontificia Universidad Cat\'{o}lica de Valpara\'{\i}so, Av. Brasil
2950, Valpara\'{\i}so, Chile }

\begin{abstract}
We studied the formation of complex patterns using a variational principle
and a standard energy functional. These patterns evolve by letting
the system to search for the optimal configuration of a high conductivity
channel, that in one dimension is equivalent to tip effect evolution
(evolution towards regions of high electric field).
\end{abstract}


\maketitle
The similarities between complex patterns produced spontaneously in
dielectric breakdown\cite{uman,dbm84,dbm95}, snowcrystals\cite{langer,rmp93,jacob93},
viscous fingers\cite{bensimon,mccloud}, etc., is suggestive for the
existence of an universal explanation for their appearance, the lack
of a general model that explains the appearance of these patterns
and the apparent difficulty of this subject, leads to the common belief
that no simple general principle can explain this diversity of complex
patterns.

Some years ago we started the search for an universal energy functional
from which complex patterns should appear. After simplifying the typical
dynamical equations for several systems, elimination of microscopic
details, and neglecting time dependent variables, we were left with
a simple Laplace equation\cite{saarlos}. 

The most general form of Laplace equation includes a conductivity
that depends on spatial coordinates, it turns out that in our model
this conductivity will have different values inside the pattern and
in the region outside the pattern. Laplace equation is a good starting
point, because this equation is a consequence of a conservation law
for the flux of the vectorial field. It is well known that Laplace
equation can be obtained from a variational principle, and the energy
functional can be found in any textbook of electromagnetism. We found
that this energy functional can produce a great variety of complex
patterns, when used in a quasi-static model that let the system to
probe different configurations for the conductivity and evolve towards
the one where the energy functional is maximized. 

In this work we studied charged capacitors that are in a static configuration
for the scalar field, but not in a state of minimal energy. For high
voltages, the dielectric is able to increase locally the permittivity
and a conducting channel can be formed between boundaries. This would
allow the system to equalize the values for the scalar field at the
boundaries, releasing the stored energy. Tip effect and screening
are essential effects in the study of these systems, we understand
our model as a generalization of tip effect evolution: instead of
evolving the channel towards the regions of high electric field, we
evolve the channel towards regions that maximize the energy functional.
Tip effect evolution and our model are equivalent for one-dimensional
systems, but for higher dimensions our model can produce complex patterns.
Other models that use tip effect, must include an independent mechanism
to produce complex patterns. Using a direct implementation of our
model and different numerical algorithms, we have obtained non-trivial
results for dielectric breakdown patterns.

To understand the underlying physics, we explain the model using an
electric scalar field $\phi$ inside a parallel plate capacitor: After
imposing boundary conditions and given the permittivity $\epsilon$,
which in general depends on coordinates, one must solve Laplace equation

\begin{equation}
\nabla\cdot(\epsilon\nabla\phi)=0,\label{laplacePhi}\end{equation}
to obtain $\phi$ in the region of interest. It is well known that
this equation can be derived, using a variational principle, from
the energy functional 

\begin{equation}
U=\frac{1}{2}\int\epsilon(\nabla\phi)dV.\label{U}\end{equation}
$U$ is also the total energy in the capacitor when the charge $Q$
is maintained constant. In the case of a parallel plate capacitor
this energy will be proportional to $d/\epsilon$, where $d$ is the
separation between the plates. Static physical systems evolve trying
to reduce the total energy, this implies a force between the plates
trying to reduce the distance $d$, and if you insert a slab of a
dielectric material having a permittivity $\epsilon$' greater than
$\epsilon$, the slab will be pulled into the capacitor. If instead
of maintaining $Q$ constant, the potential difference between the
plates $V$ is maintained constant, the energy $U$ for this system
will be proportional to $\epsilon/d$. In experiments at constant
$Q$ or $V$ there are forces trying to reduce d and to increase $\epsilon$,
then in experiments at constant $V$, $U$ is not the total energy
of the system, because the system evolves trying to increase $U$.
There is a missing energy term coming from a rearrangement of charges
in the wires to maintain $V$ constant. When this term is introduced,
the total energy is again proportional to $d/\epsilon$. The previous
discussion justifies that we study systems at constant $V$ by letting
them to evolve towards regions of higher $U$.

If V is large enough the dielectric material will break causing a
short circuit, we propose that for large V there is a new possibility
for the system to lower the total energy, a short circuit. Assuming
that $\epsilon$ can evolve locally towards bigger values, forming
a discharge channel, we could obtain complex patterns of permittivity
by letting the system to evolve towards higher values of U. Our model
is quasi-static (see below), but there are two implicit time scales
in the problem: a slow evolution for a local change in $\epsilon$,
and a fast evolution for the electric potential $\phi$ after a local
change in $\epsilon$. For these systems, complex patterns are produced
because the system is trying to lower the total energy by increasing
locally the permittivity.

Because the evolution for these systems is highly nonlinear, we have
to rely on numerical methods to study the growing of patterns. We
study these systems using a quasi-static treatment as follows: First,
set the boundary conditions which are maintained trough all the steps
in our simulation. Second, assign a fixed value $\epsilon$ to the
permittivity inside the boundaries. Third, using Laplace equation,
obtain $\phi$ after changing the permittivity locally to a greater
value $\epsilon'$, near one electrode. Fourth, find the energy U
using the new values for the permittivity and the scalar field. Fifth,
repeat steps 3 and 4 to obtain the energy values U for each of the
neighbors of one electrode. These energies are compared and the neighbor
providing the biggest energy value is added to the channel. These
steps are repeated including the new neighbors for the channel, until
a pattern develops. We used a square lattice, pattern evolution through
the diagonals is not permitted, and the pattern grows adding only
one site to the evolving pattern at each time step.

The lightning discharge in thunderstorms\cite{uman} and sparks between
charged conductors, evolve forming a branched structure that resembles
a fractal\cite{mandelbrot}. The presently accepted model of lightning\cite{dbm84,dbm95}
was developed by Niemeyer, Pietronero and Wiesmann in 1984. They include
a stochastic term that weights a probability, that is a function of
the value of the local electric field. This is known as the Dielectric
Breakdown Model (DBM) and produces a branched structure whose fractal
dimension is similar to the ones obtained experimentally for the same
geometry, Pietronero model is based on the Diffusion Limited Aggregation
(DLA) model, developed by Witten and Sander in 1981\cite{dla}.

Our results show that it is possible to obtain a branched structure
of lightning that follow from a deterministic treatment\cite{fvera1},
that only relies in minimizing the total energy in the system and
local changes in the permittivity of the medium (not the geometry
of the inner electrode) at each step of iteration. We note that our
model is almost all of the time deterministic, but for some configurations
there is degeneracy for the extreme value of the energy $U$, and
the numerical noise will be responsible for selecting the next step
in the evolution of our simulated patterns. 

We studied dielectric breakdown in the circular geometry of Fig. 1,
considering a $20\times20$ two-dimensional square lattice, where
the central point is the inner electrode and the outer electrode is
modeled as a circle. The boundary conditions, $\phi=0$ in the inner
electrode and $\phi=1$ in the outer electrode, are maintained trough
all the steps in our simulation. For each different configuration,
the numerical solution of Laplace equation was obtained using a Successive
Over Relaxation algorithm and accepted when the numerical residual
was less than $10^{-2}$, the values for the permittivity outside
the channel was $\epsilon=1$ and inside the channel was $\epsilon'=5$.
The filled boxes represent the discharge channel (sites where the
permittivity is $\epsilon'$). The circles show all possible sites
where the channel can evolve, the diameter of each of these circles
represent the value for the energy U of the system if this site is
added to the channel.

\begin{figure}
\begin{center}\includegraphics[%
  width=1.0\columnwidth]{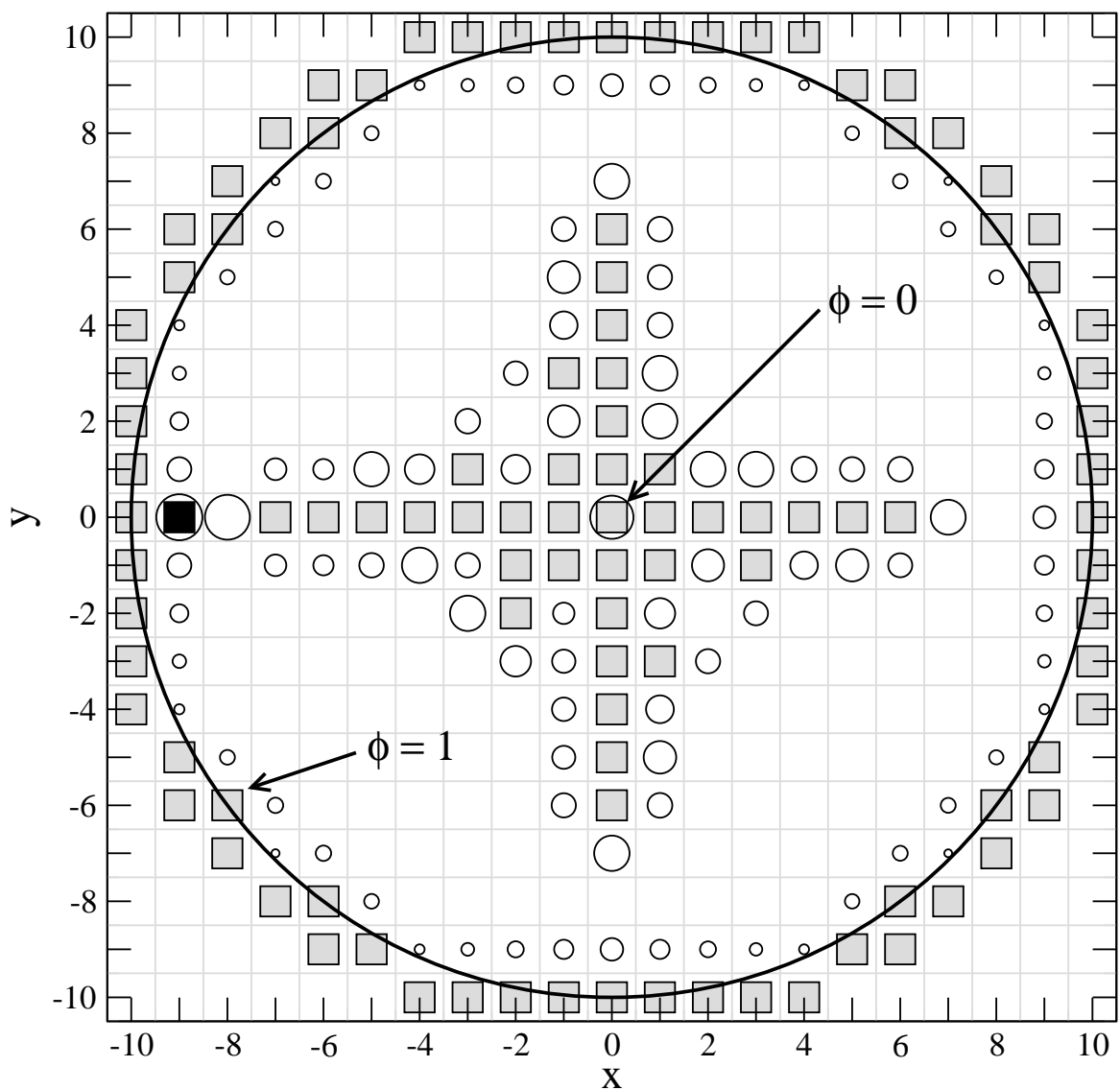}\end{center}

\caption{Discharge channel, for a $20\times20$ square lattice, showing the
attachment between the return branch (left black box) and the main
branch (inner gray boxes).}
\end{figure}

Because upward-moving discharges initiated from earth attach with
the downward-moving leader in real lightning, we considered two initial
branches: the main branch coming from the central lattice site and
the return branch coming from the outer electrode. The return branch
begin to evolve, after 36 steps of evolution of the main branch, at
the site having the black box inside (at the extreme left of the figure).
The site between this point and the main branch is the next step in
the evolution, completing the path for this discharge channel from
the inner electrode towards the outer electrode. The return branch
obtained in our numerical simulations is a highly non trivial result
and we do not know of any other work that can obtain this attachment.

Fig. 2 shows the structure of the pattern evolved from the central
lattice site for a $70\times70$ square lattice after 750 iterations.
For each different configuration, the numerical solution of Eq. 1
was obtained using a Successive Over Relaxation algorithm and accepted
when the numerical residual was less than $10^{-1}$, the values for
the permittivity outside the channel was $\epsilon=2$ and inside
the channel was $\epsilon'=6$. The evolution of the pattern shows
that opposite branches are not exactly aligned, it is possible to
find this effect in the experimental results for snowcrystal growth
shown in Fig. 3 of ref \cite{langer}. Our example also shows that
the system develops forming initially a central compact core, this
core supports the evolution of the main venous branches and diagonal
branches. We expect that secondary branches, emerging from the main
branches, would appear after some additional iterations. Because we
are using a direct implementation of the principle of least action,
that is nicely explained in the textbook of Richard Feynman\cite{feynman},
we solve Laplace equation for a permittivity change of each neighbor,
assigning permanently the value $\epsilon'$ only to the neighbor
with biggest U. This procedure is very time consuming and several
months of computing time was needed for completing this example.

\begin{figure}
\begin{center}\includegraphics[%
  width=1.0\columnwidth]{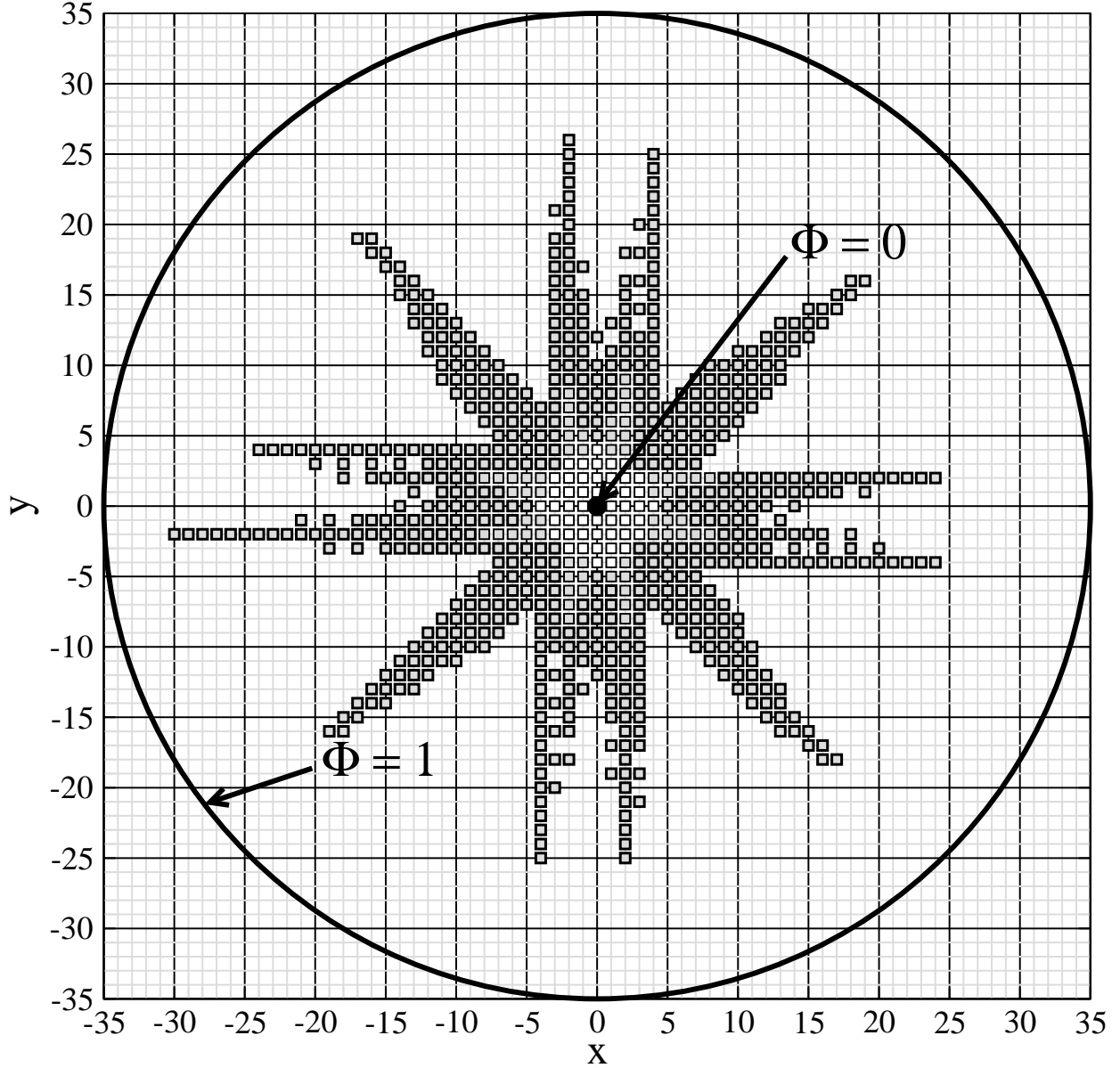}\end{center}

\caption{Pattern developed for a $70\times70$ square lattice, started at
the central site, after 750 iterations, for $\epsilon=2$ outside
the channel and $\epsilon'=6$ inside the channel.}
\end{figure}

To reduce the computing time needed for obtaining patterns, we have
re-implemented our model using adaptive grid algorithms\cite{mitchell}.
We used two grids to study the evolution of patterns: a square lattice
to evolve the pattern and an adaptive grid based on triangles to solve
Laplace equation. The adaptive grid algorithms implemented, pose the
additional restriction of solving Laplace equation in rectangular
domains. We forced the circular geometry (1/4 of a circle) by solving
the equations in the unit square, setting the permittivity to very
high values ($\epsilon=1000$), when $x^{2}+y^{2}\geq1$ and the following
boundary conditions: $\phi=1$ at the top and right boundaries, $\phi=0$
at the origin (0,0), $\phi=x$ at the bottom, and $\phi=y$ at the
left boundary.

We studied the evolution of a pattern starting from the origin using
a $100\times100$ square lattice, 20000 nodes for the adaptive grid,
$\epsilon=1$, and $\epsilon'=3$. For this sector of a circle, the
pattern begin to grow following the diagonal in a way similar to the
diagonal branches of fig. 2, after 13 steps this branch begin to depart
from the diagonal. As the number of neighbors increases, this calculation
begins to slow down.

\begin{figure}
\begin{center}\includegraphics[%
  width=1.0\columnwidth,
  keepaspectratio]{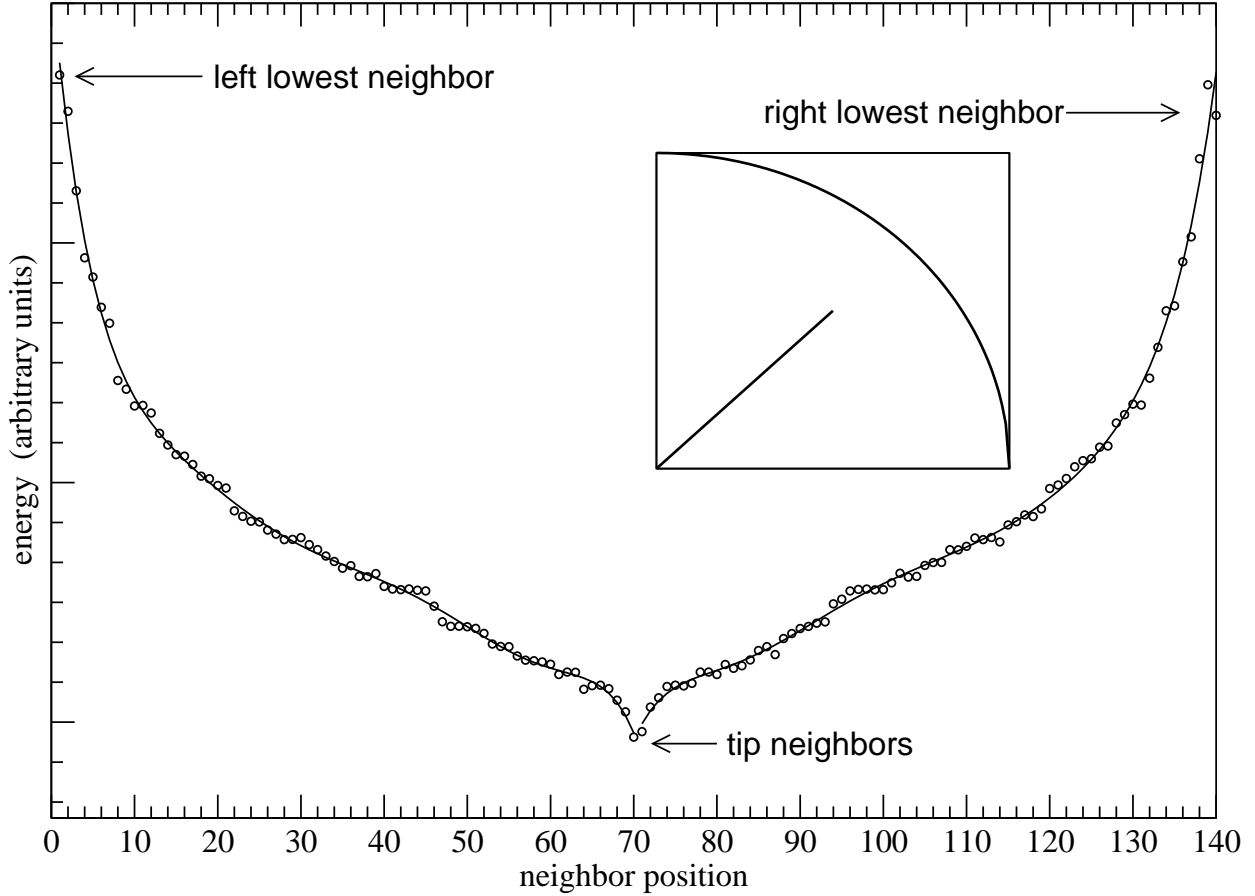}\end{center}

\caption{Energy U as a function of neighbor position for a diagonal of length
70 starting at (0,0), in a sector of a circular domain as shown in
the inset. The calculation was done in a $100\times100$ square lattice,
using 20000 nodes for the adaptive grid, $\epsilon=1$, and $\epsilon'=3$.}
\end{figure}

To gain some insight, we have carried out calculations of the energies
U for the neighbors of a given fixed pattern. We forced a one-dimensional
line of permittivity $\epsilon'$ and length 70 along the diagonal,
starting at (0,0), as shown in the inset of figure 3. In this figure
we plot the energies U obtained after a local change of each neighbor
permittivity towards $\epsilon'$, neighbors near position 1 correspond
to the line left lowest neighbors, neighbors near position 70 correspond
to the line tip neighbors, and neighbors near position 140 correspond
to the line right lowest neighbors. In this plot we fitted 7th grade
polynomials for the left and right branches. Because the system evolves
towards high values of U, this line would evolve towards the lowest
neighbors and not towards the tip neighbors. This evolution is contrary
to tip effect evolution and is a consequence of the finite value $\epsilon'$
and geometry.

To investigate the effect of geometry, we studied the evolution of
a pattern starting from the central site of the bottom boundary in
a unit square domain and the following boundary conditions: $\phi=1$
at the top, $\phi=0$ at the bottom, $\phi=y$ at the left boundary,
and $\phi=y$ at the right boundary. Using a $100\times100$ square
lattice to evolve the pattern, 20000 nodes for the adaptive grid,
$\epsilon=1$, and $\epsilon'=20$, the pattern grows as a vertical
one-dimensional line. This is exactly what is expected from tip effect
evolution and shows that the numerical error is well controlled in
our simulations. 

To show the differences between our model and tip effect evolution,
we forced a 1-dimensional vertical line of permittivity $\epsilon'$
and length 70, starting at the central site of the bottom boundary,
as shown in the inset of fig. 4. In this figure, we plot the energies
U obtained after a local change of each neighbor permittivity towards
$\epsilon'$, as a function of neighbor position. We used the same
boundary conditions mentioned in the previous paragraph, $\epsilon=1$
and $\epsilon'=3$. In this case, a permittivity change in the neighbor
just above the tip gives the pattern of highest U, explaining the
previous vertical line evolution. This plot provides much more information
than just tip effect evolution. If an independent mechanism, like
surface tension, is introduced to prevent tip evolution, the left
and right lowest neighbors could begin to evolve. 

\begin{figure}
\begin{center}\includegraphics[%
  width=1.0\columnwidth]{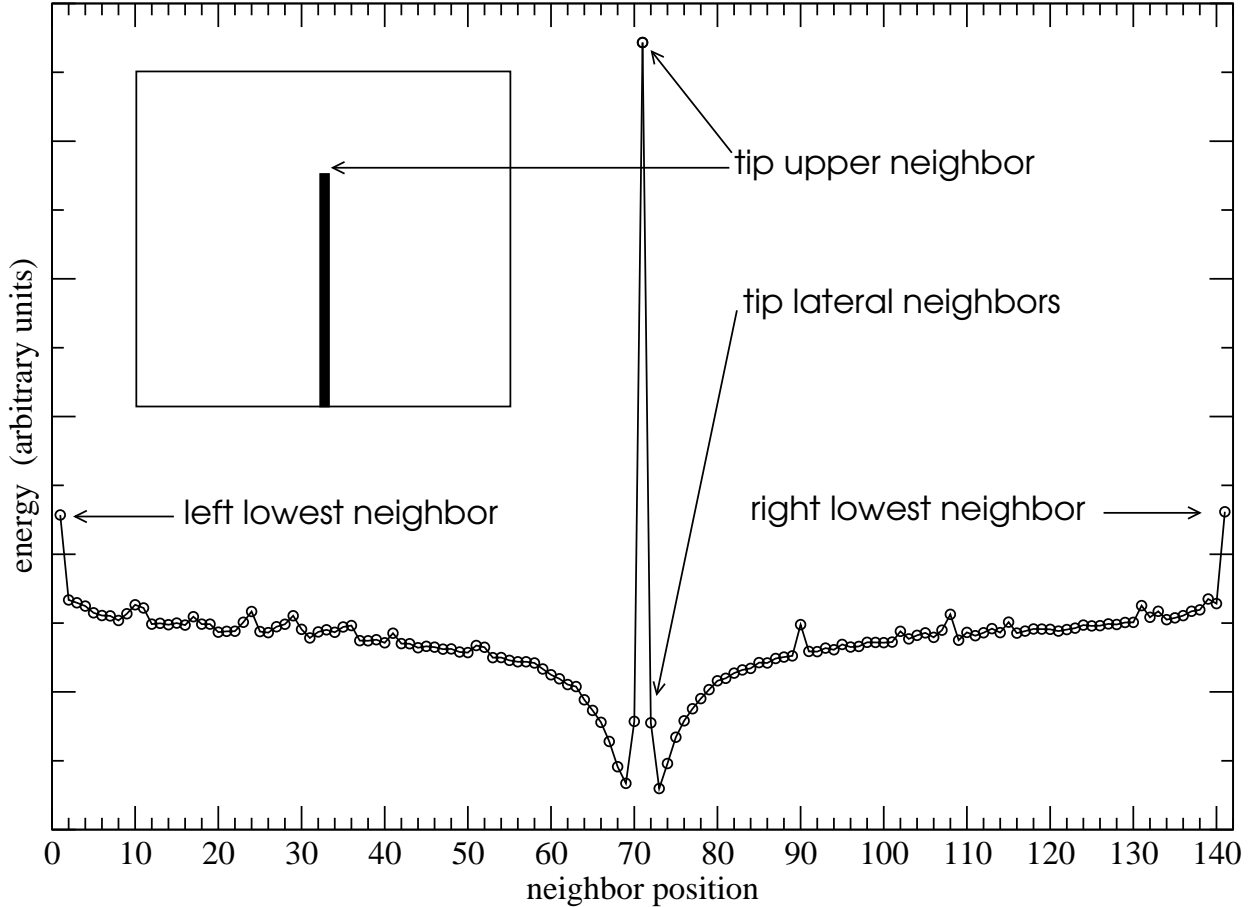}\end{center}

\caption{Energy U as a function of neighbor position for a vertical line of
length 70 starting at the center of the bottom boundary, in a square
domain. The calculation was done in a $100\times100$ square lattice,
using 20000 nodes for the adaptive grid, $\epsilon=1$, and $\epsilon'=3$.}
\end{figure}

To summarize: We have obtained non trivial patterns using a simple
and well known energy functional. Contrary to other models, our model
is not stochastic and the generated patterns are not of dynamical
origin. Our model needs much more computing time than other models
in the literature, because the system has to probe different possible
configurations and select the one minimizing the total energy. We
have seen the appearance of a return branch, which is typically found
in real lightning. We obtained the breaking of chiral symmetry in
a global pattern, as a consequence of degeneracy and time evolution.
In circular geometries, the patterns obtained using our model, show
branches that are roughly identical in length, we obtained this very
important result straightforward by letting the system to evolve locally
towards a configuration of minimal energy. For the same set of parameters
we have seen a pattern to grow first in a compact form, and after
reaching a critical size the system begin to form branches. For some
geometries tip growing is favored, for other geometries the system
try to increase the width of the channel. Because the evolving pattern
changes the geometry, and the evolution depends on system history,
there are many possibilities for complex patterns to appear. In the
near future we expect to obtain fractal like structures from our numerical
simulations. 

We hope that this mechanism of reducing energy by increasing conductivities
could be applied to many others different pattern forming systems.

\end{document}